\documentclass[12pt]{article}
\newcommand{\be}{\begin{equation}}
\newcommand{\ee}{\end{equation}}
\newcommand{\bea}{\begin{eqnarray}}
\newcommand{\eea}{\end{eqnarray}}
\title{\Large Comparison of Nonrelativistic limit of Amelino-Camelia and MS Doubly Special Relativity
 \author{Nosratollah Jafari\thanks{Email: \texttt{nosrat.jafari@gmail.com}
 } , Mostafa PourNemati\\  \small \textit{Department of Physics,
 Semnan University, Semnan, Iran}}}

\begin{document}

\maketitle

\begin{abstract}

This paper is devoted to the study of the nonrelativitic limit of
Amelino-Camelia Doubly Special Relativity, and the corresponding
modified Klein-Gordon and Dirac equations. We show that these
equations reduce to the Schr\"{o}dinger equations for the particle
and the antiparticle with different inertial masses. However,
their rest masses are the same. M. Coraddu and S. Mignemi have
studied recently the non relativistic limit of the Magueijo-Smolin
Doubly Special Relativity. We compare their results with our
study, and show that these two models are reciprocal to each other
in the nonrelativitic limit. The different inertial masses also
leads to the CPT violation.

 \end{abstract}

\vspace{5cm}


\newpage

\section{Introduction}
Doubly Special Relativity (DSR) theories have been proposed ten
years ago for the nonlinear modification of Special Relativity.
These theories have two invariant scales, the speed of light $c$
and the Planck energy $E_{p}=\sqrt{\hbar c^5/G}\simeq 10^{19}$
GeV. Magueijo-Smolin (MS) DSR \cite{MS} and Amelino-Camelia DSR
\cite{Am1, Am2} are the two main examples of these theories.
Although, these model have similar structure, they belong to
different realizations of kappa-Pioncare algebras \cite{Kowal}.

According to the special relativity and relativistic quantum
mechanics the Schr\"{o}dinger equation is the nonrelativistic
limit of the Klein-Gordon and Dirac equations \cite{Gross}. Also,
we have the same mass for the particle and the antiparticle, and
this fact can be seen easily from the dispersion relation
$E=\pm\sqrt{m^2+p^2}$.

The dispersion relation has been modified in the DSR theories,
which leads to the modified Klein-Gordon and Dirac equations. It
will be interesting to study the nonrelativistic limit of these
modified equations. The nonlinearity of the dispersion relation
and the fact that the dispersion relation is not invariant under
space inversion and time reversal also imply the violation of CPT
invariance.

Recently, M. Coraddu and S. Mignemi have studied the
nonrelativitic limit of the MS DSR and the corresponding modified
Klein-Gordon and Dirac equations \cite{Mig}. They illustrated that
the particle and the antiparticle rest masses are different.
However, their inertial masses are the same. Besides, they have
showed that the modified Klein-Gordon and Dirac Equations in the
MS model reproduce nonrelativistic quantum mechanics. To continue
their proposal we want to study the nonrelativistic limit of
Amelino-Camelia Doubly Special Relativity and the corresponding
modified Klein-Gordon and Dirac equations.

We show that the corresponding modified Klein-Gordon and Dirac
equations reduce the Schr\"{o}dinger equations for the particle
and the antiparticle with different inertial masses. The
difference between these two masses is proportional to
$mc^2/E_{p}$ in the first order of approximation. However, their
rest masses are the same. Different inertial masses also leads to
the violation of CPT invariance.

We compare the nonrelativistic limit of MS and Amelino-Camelia DSR.
Our results are reciprocal to the M. Coraddu and S. Mignemi results.
 We can interpret MS DSR as modifying rest mass \cite{Mig}, and
 Amelino-Camelia DSR as modifying momentum.

In the following section, we summarize some basics of Amelino-Camelia DSR.
 In section 3, we study the non-relativistic limit of modified
dispersion relation in Amelinio-Camelia DSR, and make some
comparisons with the nonrelativistic limit of MS DSR.
The non-relativistic limits of ordinary and modified Klein-Gordon
and Dirac equations is presented in section 4. Then, in section 5
 we have some conclusions.

\section{Amelino-Camelia DSR}
\par The essence of Amelino-Camelia DSR is the modification of
ordinary special relativistic boosts

 $$ \textit{K}_a=ip_a\frac{\partial}{\partial{E}}+
 iE\frac{\partial}{\partial{p_a}}, $$
to preseve the following nonlinear dispersion relation invariant
\cite{Am2}. This modified dispersion relation is
  \be\label{e:1} 2k^2[\hspace{1mm}\cosh(\frac{E}{k})-
\cosh(\frac{m}{k})\hspace{1mm}]=\textbf{p}^2 e^{E/k}, \ee
in which $k$ is the Planck energy $E_p$ and $a$ is an index varies
 from 1 to 3. He proposed the modified boosts
\be \textit{B}_a=ip_a\frac{\partial}{\partial{E}}+
i(\hspace{1mm}\frac{1}{2k}\textbf{p}^2
    +k\frac{1-e^{-2E/k}}{2} \hspace{1mm})\frac{\partial}{\partial{p_a}}
     -i\frac{p_a}{k}(p_b \frac{\partial}{\partial{p_b}}),
     \ee
which leave invariant the mentioned dispersion relation. Also, we
can find new transformations for this doubly special relativity
which is different form the Lorentz transformations \cite{Am3}.

The modified Klein-Gordon equation is obtained from (\ref{e:1}) by
substituting differential operators for
$E=i\frac{\partial}{\partial t}$ and $\textbf{p}=-i\vec{\nabla} $
 in the fashion standard in quantum mechanics  \be\label{e:2} \left[\nabla^2
-2k^2\exp{(\frac{-i}{k}\frac{\partial}{\partial
t})}\left(\hspace{1mm}\cosh(\frac{-i}{k}\frac{\partial}{\partial
t})-\cosh(\frac{m}{k})\hspace{1mm}\right)\right]\Psi(\vec{x},t)=0.
\ee Also, we can construct deformed Dirac equation as

\be \label{e:3} \left( \gamma^\mu \textit{D}_\mu -I
\right)\Psi(\textbf{p})=0 \ee where $\textit{D}_\mu$ is the modified
dirac operator

$$ \textit{D}_0=\frac{e^{E/k}-\cosh(m/k)}{\sinh(m/k)} \hspace{1mm}, $$

$$ \textit{D}_a=\frac{p_a}{p}\frac{\left(2e^{E/k}[\cosh(E/k)-\cosh(m/k)]\right)^{1/2}}{\sinh(m/k)} \hspace{1mm}, $$
and $\gamma^\mu$ are the familiar Dirac $\gamma$ matrices
\cite{Am2}.

\section{Nonrelativistic limit}
\par In the ordinary special relativity, we go to the classical
non-relativistic limit for a free particle by expanding ordinary
dispersion relation

\be E^2- c^2 \textbf{p}^2 =m^2c^4   \ee in
$(\textbf{p}^2c^2/m^2c^4) \ll 1$ limit, which leads to

\be E= \pm\sqrt{ \textbf{p}^2c^2 + m^2c^4 }\simeq \pm(mc^2
+\frac{\textbf{p}^2}{2m}+...)\hspace{1mm}. \ee The first term on
the right hand side is the rest mass energy and the second term is
the classical kinetic energy.

In Amelino-Camelia DSR, we expand modifed dispersion relation
(\ref{e:1}) in $O(E^3/k^3)$ approximation and we obtain

\be \label{e:12}
E^2-\textbf{p}^2c^2-\frac{\textbf{p}^2c^2}{k}E=m^2c^4.\ee Solving
this equation as a second order equation for $E$, we have

$$ E=\frac{\textbf{p}^2c^2/k\pm \sqrt{\textbf{p}^4c^4/k^2+4(\textbf{p}^2c^2+m^2c^4)}}{2}.  $$
In the $(\textbf{p}^2c^2/m^2c^4) \ll 1$ limit we have

$$ E\simeq \frac{\textbf{p}^2c^2}{2k}\pm mc^2(1+\frac{\textbf{p}^2}{2m^2c^2}).$$
The positive sign corresponds to the classical non-relativistic
limit \be \label{e:11}
 E=m c^2 +\frac{\textbf{p}^2}{2m^{+}},\ee in which we
assume \be\label{e:8} m^{+}=\frac{m}{1+\frac{mc^2}{k}}.\ee
Thus, we find that the particle inertial mass has been modified
 by amount of $mc^2/k$, but the rest mass of particle remains
unmodified. By inertial mass  we mean the mass which appears in
dominator of $\textbf{p}^2/2m^{+}$.
Also, by defining \be
\label{e:9} m^{-}=\frac{m}{1-\frac{mc^2}{k}}\ee we can interpret
the negative sign solution as an antiparticle \be
E=-mc^2-\frac{\textbf{p}^2}{2m^{-}},\ee moving in the opposite
direction of time with modified inertial mass $m^{-}$.

Different inertial mass $m^{+}$ and $m^{-}$ in the equations
(\ref{e:8}) and (\ref{e:9}) leads to the violation of CPT
invariance. However, the interpretation of the inertial mass as the
mass of an elementary particle may have some difficulties. In principle,
 modification of mass-shell relation Eq.(\ref{e:12}) can be
interpreted as the violation of CPT.

We can compare this case with nonrelativistic limit of the MS
model \cite{Mig}. They reached to the modified relation ($ E=m^{+}
c^2 +\frac{\textbf{p}^2}{2m}$) instead of Eq.(\ref{e:11}) in which
$m^{+}$ has the same value with Eq.(\ref{e:8}). In this model the rest mass
of the particle is modified, and the inertial mass remains the
same. Thus we have an interesting result, that the nonrelativistic
limit of Magueijo-Smolin DSR is reciprocal to the nonrelativistic limit of Amelino-Camelia DSR.

Also, we can calculate the non-relativistic limit of the velocity
of a particle. The group velocity is \be
\textbf{v}_g=\frac{\partial E}{\partial \textbf{p}}\simeq
\frac{\textbf{p}}{m^{+}}, \ee and the particle velocity is \be
\textbf{v}_{particle}=\frac{
\textbf{p}c^2}{E}\simeq\frac{\textbf{p}}{m}. \ee In the ordinary
special relativity we have the same value of $\textbf{p}/m $ for
the group velocity $\textbf{v}_{g}$ and the particle velocity
$\textbf{v}_{particle}$. Here, we showed that the group velocity
differ from the special relativistic value but the particle
velocity is the same as in the special relativity. In the
nonrelativistic limit of MS DSR the situation for velocities are
inverse to our case and we have $ \textbf{v}_g\simeq \textbf{p}/m
$ and $\textbf{v}_{particle}\simeq \textbf{p}/m^{+}$ \cite{Mig}.

We summarize comparison of the nonrelativistic limit of
Amelino-Camelia and MS doubly special relativity in the following
table.
$$ \begin{tabular}{|@{$ $} c|cc |}\hline
Amelino-Camelia DSR&Magueijo-Smolin DSR &\\\hline $ E=m c^2
+\textbf{p}^2/2m^{+} $& $ E=m^{+} c^2+\textbf{p}^2/2m $&\\\hline
$\textbf{v}_g=\textbf{p}/m^{+}
$&$\textbf{v}_g=\textbf{p}/m$&\\\hline$\textbf{v}_{particle}=\textbf{p}/m
$&$\textbf{v}_{particle}=\textbf{p}/m^{+} $&\\\hline

\end{tabular} $$

\section{Nonrelativistic limit of the modified Klein-Gordon and Dirac equations}

We now consider the nonrelativistic limit of  modified
Klein-Gordon and Dirac equations. The nonrelativistic
limit of the ordinary Klein-Gordon and Dirac equations can be found in many
relativistic quantum mechanics and quantum field theory books
\cite{Gross}.
\\\\
\textbf{4.1- Ordinary Klein-Gordon Equation:} Ordinary
Klein-Gordon equation reads
$$(\frac{1}{c^2}\frac{\partial^2}{\partial
t^2}-\vec{\nabla}^2+m^2c^2)\Psi(\vec{x},t)=0.$$ We define new
operator $M=\sqrt{m^2c^4-c^2\vec{\nabla^2}}$ and rewrite
 the Klein-Gordon equation as
 $$-\frac{\partial^2}{\partial t^2}\Psi=M^2\Psi. $$
By introducing new fields
$$ \phi^{\pm}=\Psi \pm iM^{-1}\frac{\partial \Psi}{\partial
t} ,$$ we can convert the Klein-Gordon equation to the first order
equations
$$ i\frac{\partial \phi^{\pm}}{\partial t}= \pm M\phi^{\pm}.$$

In the non-relativistic limit $(\textbf{p}^2c^2/m^2c^4) \ll 1$ or
large $c$ we have
$$M\simeq mc^2-\frac{1}{2m}\vec{\nabla^2}.$$
For subtracting the rest mass energy, we transform to the fields
$\tilde{\phi}^{\pm}=\exp{(\pm imc^2t)}\phi^{\pm} $ and we reach to
the Schr\"{o}dinger equations
$$ i\frac{\partial \tilde{\phi}^{\pm}}{\partial t}= \mp \frac{1}{2m}\vec{\nabla}^2\tilde{\phi}^{\pm}.$$
The positive sign in the $\tilde{\phi}^{\pm}$ corresponds to the
particle solution and the negative sign to the antiparticle case
\cite{Mig}.
\\\\
  \textbf{4.2- Modified Klein-Gordon Equation:} Modified Klein-Gordon equation Eq.($\ref{e:2}$) in \textit{O}($E^3/k^3$)
  approximation is

 $$ -\frac{1}{c^2}\frac{\partial^2}{\partial t^2}\Phi+ \frac{i}{k}\frac{\partial}{\partial
  t}\vec{\nabla}^2\Phi-(m^2c^2-\vec{\nabla}^2)\Phi=0. $$ This equation can be
  rewritten as
  $$ -\frac{\partial^2}{\partial t^2}\Phi=\left[m^2c^4-\left(1+\frac{i}{k}\frac{\partial}{\partial
  t}\right)\vec{\nabla}^2 \right]\Phi. $$
By defining new operator
 $$\tilde{M}=\sqrt{m^2c^4-c^2\left(1+\frac{i}{k}\frac{\partial}{\partial
t}\right)\vec{\nabla^2}},$$ and introducing new fields $$
\psi^{\pm}=\Phi \pm i\tilde{M}^{-1}\frac{\partial \Phi}{\partial
t} .$$ We can convert the modified Klein-Gordon equation to the
first order equations
$$ i\frac{\partial \psi^{\pm}}{\partial t}= \pm \tilde{M}\psi^{\pm}.$$

In the non-relativistic limit $(\textbf{p}^2c^2/m^2c^4) \ll 1$ or
large $c$ we have
$$\tilde{M}\simeq mc^2-\frac{1}{2m}\left(1+\frac{i}{k}\frac{\partial}{\partial
t}\right)\vec{\nabla^2}.$$ By transforming to the fields
$\tilde{\psi}^{\pm}=e^{\pm imc^2t}\psi^{\pm} $, we reach to the
Schr\"{o}dinger equations

$$ i\frac{\partial \tilde{\psi}^{\pm}}{\partial t}= \mp \frac{1}{2m^{\pm}}\vec{\nabla}^2\tilde{\psi}^{\pm}.$$
As expected, the particle and the antiparticle states satisfy in
the Schr\"{o}dinger equations with the modified inertial masses
$m^{+}$ and $m^{-}$.
\\\\
\textbf{4.3- Dirac Equation:} Ordinary Dirac Equation is
 $$ ( i\gamma^{\mu} \frac{\partial}{\partial x^{\mu}}-m)\psi=0. $$  Defining
 two-component form as $$ \psi(\vec{x},t)=\left(\begin{array}[c]{ll}\chi\\ \eta
\end{array}\right)$$
and using standard form of $\gamma$ matrices
$$\gamma^0=\left(\begin{array}[c]{ll} 1 &\hspace{3mm} 0\\0&-1
\end{array}\right),\hspace{10mm}
 \gamma^i=\left(\begin{array}[c]{ll}\hspace{2mm} 0 &\hspace{1mm} \sigma_{i}\\-\sigma_{i}&\hspace{1mm}0
\end{array}\right)
$$
we reach to coupled equations \be \label{e:10}
\left\{\begin{array}{cl}(E-m)\hspace{1mm}\chi=\vec{\sigma}.\vec{p}\hspace{1mm}
\eta
 \vspace{4mm}
 \\
  (E+m)\hspace{1mm}\eta=\vec{\sigma}.\vec{p}\hspace{1mm}\chi
  \end{array}\right. \ee
Note that $p^{\mu}=(E,\textbf{p})$, $p_{\mu}=(E,-\textbf{p})$ and
$\eta_{\mu \nu}=\textrm{diag}(1,-1,-1,-1)$.

In the nonrelativistic limit for the particle solution, we take
$E\simeq m $ and $(\textbf{p}^2c^2/m^2c^4) \ll 1$ which leads to
$\eta\simeq (\vec{\sigma}.\vec{p}/2m) \chi$. Substituting this in
Eq.(\ref{e:10}) and introducing $\chi'=e^{+imt}\chi$, we reach to
the Schr\"{o}dinger equation
  \be i\frac{\partial \chi'}{\partial
  t}\simeq-\frac{1}{2m}\vec{\nabla}^2 \chi'. \ee
For the antiparticle solution in the nonrelativistic limit we take
$E\simeq -m $. Doing as in particle
  case and introducing $\eta'=e^{-imt}\eta$, we can show that also $\eta'$ satisfies in a similar
Schr\"{o}dinger equation
  \be i\frac{\partial \eta'}{\partial
  t}\simeq+\frac{1}{2m}\vec{\nabla}^2 \eta'. \ee
\\
  \textbf{4.4- Modified Dirac Equation:} Modified dirac equation Eq.($\ref{e:3}$) in the \textit{O}($E^2/k^2$)
  approximation is $$ \left[ i\gamma^{0}\frac{1}{c}\frac{\partial}{\partial
t}+i\gamma^{i}\frac{\partial}{\partial x^i}
 \left(1+\frac{i}{2k}\frac{\partial}{\partial t}\right) -m \right]\tilde{\psi}=0. $$
In this case, we obtain the two-component by introducing
$$\tilde{\psi}(\vec{x},t)=\left(\begin{array}[c]{ll}\tilde{\chi}\\ \tilde{\eta}
\end{array}\right)$$
Doing as in the unmodified case, we reach to the following coupled
equations
$$ \label{e:6}
\left\{\begin{array}{cl}\left[E-m\right]\hspace{1mm}\tilde{\chi}=(1+\frac{E}{2k})\vec{\sigma}.\vec{p}\hspace{1mm}
\tilde{\eta}
 \vspace{4mm}
 \\
  \left[E+m\right]\hspace{1mm}\tilde{\eta}=(1+\frac{E}{2k})\vec{\sigma}.\vec{p}\hspace{1mm}\tilde{\chi}
  \end{array}\right. $$
In  the nonrelativistic limit $(\textbf{p}^2c^2/m^2c^4) \ll 1$, we
take $E\simeq m $ for the particle solution and we have
$$\tilde{\eta}\simeq(1+\frac{E}{2k})\frac{\vec{\sigma}.\vec{p}}{2m}\tilde{ \chi}.$$
By introducing $\tilde{\chi'}=\exp{(+imc^2t)}\tilde{\chi}$, we
reach to the following Schr\"{o}dinger equation

  \be i\frac{\partial \tilde{\chi'}}{\partial
  t}\simeq-\frac{1}{2m^{+}}\vec{\nabla}^2 \tilde{\chi'}. \ee
 We note that $\tilde{\chi'}$ satisfies in the
Schr\"{o}dinger equation with the modified inertial mass $m^{+}$.

Also, for the antiparticle we take $E\simeq -m $ and we have
$$\tilde{\chi}\simeq-(1+\frac{E}{2k})\frac{\vec{\sigma}.\vec{p}}{2m}\tilde{ \eta}.$$
 If we define $\tilde{\eta'}=\exp({-im c^2 t})\tilde{\eta}$ we can show that $\tilde{\eta'}$ satisfies in a similar
Schr\"{o}dinger equation
  \be i\frac{\partial \tilde{\eta'}}{\partial
  t}\simeq+\frac{1}{2m^{-}}\vec{\nabla}^2 \tilde{\eta'}, \ee
in which the modified mass $m^{-}$ is given by Eq.($\ref{e:9}$)

\section{Conclusion and some remarks}

In this paper, we showed that the nonrelativistic limit of
Amelino-Camelia DSR leads to the particle and the antiparticle
with different inertial masses $m^{+}$ and $m^{-}$ but the same
rest mass. We interpreted this difference as a sign of the CPT
violation.

Also, we studied the corresponding modified Klein-Gordon and Dirac
equations, which reproduce the Schr\"{o}dinger equations in the
nonrelativistic limit with these modified masses for the particle
and the antiparticle.

The ratio $|m^{+}-m^{-}|/m$ is equal to $2mc^2/k$. We can use this
ratio to find a lower bound on the amount of $k$. In DSR theories
$k$ is a fundamental constat with dimension of energy and in
principle can be different form the Planck energy $E_p=\sqrt{\hbar
c^5/G}\simeq 10^{19}$.

We compared the nonrelativistic limit of Amelino-Camelia and
Magueijo-Smolin DSR. In this limit we reached to the $E=m c^2
+\textbf{p}^2/2m^{+} $ for Amelino-Camelia DSR, also we $ E=m^{+}
c^2 +\textbf{p}^2/2m $ for MS DSR in this limit \cite{Mig}. These
are seem to be the natural results of the dispersion
 relations. If we put the \textit{O}($E^3/k^3$) approximation of
 Amelino-Camelia dispersion relation Eq.(\ref{e:1}) and the MS dispersion
 relation \cite{MS}
 together:

$$
\left\{\begin{array}{cl}E^2=
\textbf{p}^2c^2+m^2c^4(1-E/k)^2\hspace{1cm}\textrm{Magueijo-Smolin}
 \vspace{4mm}
 \\E^2\simeq \textbf{p}^2c^2(1+E/k)+m^2c^4 \hspace{1cm} \textrm{Amelino-Camelia}
  \end{array}\right. $$
Then the observation of their similarities and differences is
easy, and the interpretation of the nonrelativistic limit of MS
DSR as modifying rest mass and Amelino-Camelia DSR as modifying
momentum is not difficult. Thus, in this limit the MS and
Amelino-Camelia DRS theories seems to be complementary to each
other.

\section{ Acknowledgment}
We want to thanks to Dr. Ahmad Shariati for his kindly and useful
discussions.

\end{document}